\documentclass[%
 reprint,
 amsmath,amssymb,
 aps,
]{revtex4-2}

\usepackage{graphicx}
\usepackage{dcolumn}
\usepackage{bm}
\usepackage{hyperref}
\usepackage{newtxtext}
\usepackage{newtxmath}
\usepackage{titlesec}
\usepackage{hyperref}
\usepackage{float}
\usepackage{subcaption}
\usepackage{verbatim}

\newcommand{\xsub}[1]{%
  \mbox{\scriptsize\begin{tabular}{@{}c@{}}#1\end{tabular}}%
  }

\begin{document}

\preprint{APS/123-QED}

\title{Unstable magnetic reconnection self-generates turbulence}
\author{Nick Williams}
 \email{nicholas.williams-2@manchester.ac.uk}
 \altaffiliation{Department of Mechanical and Aerospace Engineering \\ The University of Manchester}
\author{Alessandro De Rosis}%
 \email{alessandro.derosis@manchester.ac.uk}
 \altaffiliation{Department of Mechanical and Aerospace Engineering \\ The University of Manchester}
\author{Alex Skillen}
 \email{alex.skillen@manchester.ac.uk}
 \altaffiliation{Department of Mechanical and Aerospace Engineering \\ The University of Manchester}

\date{\today}

\begin{abstract} 
Magnetic reconnection and turbulence are deeply intertwined in magnetohydrodynamic flows, yet how reconnection self-generates turbulence remains unclear. Using an ensemble of high-resolution three-dimensional direct numerical simulations of an unstable magnetised jet with an initially weak mean field, we demonstrate a self-sustained transition from laminar reconnection to fully developed turbulence in the absence of external forcing. We show that a three-dimensional current-sheet instability triggers stochastic reconnection, leading to persistent turbulent energy injection. Energy-budget analysis reveals that the coupling between the turbulent electromotive force and the magnetic mean shear dominates turbulent production, with magnetic fluctuations subsequently transferring energy to the kinetic field through a nonlinear cascade.
\end{abstract}

\maketitle


\textit{Introduction -} Magnetohydrodynamic (MHD) turbulence occurs in electrically conducting fluids permeated by magnetic fields and plays a central role in a broad range of astrophysical and fusion plasmas, such as the solar wind \cite{bruno_solar_2013}, interstellar medium \cite{falceta-goncalves_turbulence_2014, burkhart_diagnosing_2021}, and magnetically confined plasmas \cite{ida_experimental_2025, morales_magnetohydrodynamically_2015}. These environments are typically characterised by large Reynolds and magnetic Reynolds numbers, rendering turbulence an inherent and persistent feature of their evolution.\\
\indent A central source of the complexity in MHD turbulence is magnetic reconnection~\cite{ji_magnetic_2022}. During reconnection, magnetic field lines of opposing polarity rapidly realign, converting magnetic energy into current sheets and accelerating plasma through enhanced Lorentz forces. This mechanism enables rapid magnetic energy release and redistribution across scales~\cite{pontin_magnetic_2022}.\\
\indent 
Notably, reconnection is not confined to the inertial range of pre-existing turbulence. Large-scale events, such as coronal mass ejections~\cite{webb_coronal_2012}, involve the reconnection of magnetic structures on scales far exceeding those of local turbulent fluctuations~\cite{marino_scaling_2023}. This distinction motivates the separation between `turbulence-driven reconnection', in which reconnection is triggered by interactions within an established turbulent cascade, and `reconnection-driven turbulence', in which large-scale reconnection injects energy directly into turbulent motions~\cite{schekochihin_mhd_2022}.\\
\indent Despite substantial progress in characterising reconnection rates~\cite{lazarian_3d_2020, grehan_comparison_2025, loureiro_nonlinear_2020, vicentin_investigating_2025}, current sheet topologies~\cite{ browning_kink_2024,hosking_forced_2025,kalluri_quantifying_2026, russell_flux-rope-mediated_2025}, the spectral and anisotropic properties of MHD turbulence~\cite{brandenburg_inverse_2025, hosking_reconnection-controlled_2021, milanese_dynamic_2020, oughton_critical_2020, pouquet_helicity_2019, zhang_resistive_2026}, and mean-field-fluctuation coupling~\cite{bian_decoupled_2019, arro_spatiotemporal_2025}, the fundamental mechanics of the reconnection-driven turbulence cascade remain unclear. Although prior studies  have examined the role of the fluctuating electromotive force (EMF) the exact pathways for the production and redistribution of energy in reconnection-driven turbulence are still not fully understood~\cite{huang_turbulent_2016, oishi_self-generated_2015, beg_evolution_2022}. \\
\indent In this Letter we demonstrate that the primary mechanism underlying the self-generation of turbulence is the production of Alfvén waves through the interaction between the turbulent electromotive stresses and the mean magnetic shear, while kinetic turbulence arising in reconnection outflows plays a secondary, though important, role. We isolate the mechanisms for the transition of pseudo-two-dimensional Sweet-Parker reconnection~\cite{parker_sweets_1957, sweet_neutral_1958}  into a regime of self-generated and self-sustaining turbulent reconnection (SGTR) regions~\cite{beg_evolution_2022}, followed by the onset of stochastic three dimensional (3D) reconnection and ultimately a fully turbulent flow. Our results complement the Lazarian \& Vishniac (LV99) perspective of stochastic 3D reconnection~\cite{lazarian_reconnection_1999}, clarifying the energy-production pathways that link reconnection dynamics to turbulence. 

\indent \textit{Methods -} We consider an incompressible, isothermal magnetohydrodynamic (MHD) fluid in a three-dimensional (3D) Cartesian coordinate system $x_1\!-\!x_2\!-\!x_3$. The velocity and magnetic fields are $\boldsymbol{u}(\boldsymbol{x},t)$ and $\boldsymbol{b}(\boldsymbol{x},t)$, with components $u_i$ and $b_i$, where $\boldsymbol{x}$ denotes the position vector and $t$ the time. The governing equations are
\begin{eqnarray}
\partial_t u_i + u_j \partial_j u_i = -\partial_i p &+& \nu \partial_{jj} u_i + b_j \partial_j b_i - \partial_i \frac{\|\boldsymbol{b} \|^2}{2}, \label{eq:NS}\\
\partial_t b_i + u_j \partial_j b_i - b_j \partial_j u_i &=& \eta \partial_{jj} b_i, \label{eq:induction}
\end{eqnarray}
subject to the solenoidal constraints
\begin{equation}
\partial_i u_i = 0,
\qquad
\partial_i b_i = 0.
\label{eq:divfree}
\end{equation}
Here $p$ denotes the hydrodynamic pressure, $\nu$ the kinematic viscosity, and
$\eta$ the magnetic diffusivity. Indices $i,j=1,2,3$ span Cartesian components, repeated indices imply summation, and $\partial_j \equiv \partial/\partial x_j$. \\
\indent Equations~(\ref{eq:NS})--(\ref{eq:divfree}) are solved using an in-house three-dimensional direct numerical simulation code. Spatial derivatives are discretised using a sixth-order central finite-difference scheme, while the pressure Poisson equation is solved using a spectral method. Time integration is performed with a fourth-order explicit Runge--Kutta scheme. Numerical violations of the divergence-free constraint on the magnetic field are controlled using a divergence-cleaning procedure~\cite{brackbill_effect_1980}, ensuring $\partial_ib_i$ remains below $10^{-13}$.\\
\indent Following Mak \emph{et al.}~\cite{mak_vortex_2017}, we simulate an unstable, decaying magnetised jet. The initial velocity and magnetic fields are 
\begin{eqnarray}
\boldsymbol{u}(\boldsymbol{x},t=0)
&=&
\begin{pmatrix}
\mathrm{sech}^2(x_2)\\
0 \\
0
\end{pmatrix}
+
\begin{pmatrix}
10^{-3}\sin(\sigma x_1) + \alpha \\
10^{-3}\sin(\sigma x_1) + \alpha \\
\alpha
\end{pmatrix}
\mathrm{e}^{-x_2^2},\\
\boldsymbol{b}(\boldsymbol{x},t=0)
&=&
\begin{pmatrix}
0.015\\
0 \\
0
\end{pmatrix}
\label{eq:perturbation}
\end{eqnarray}
where $\sigma = 0.9$ corresponds to an unstable mode and $\alpha(\mathbf{x}) \sim \mathcal{U}(-10^{-5}, 10^{-5})$ is a uniformly distributed random field representing small-scale background fluctuations. The computational domain is a periodic box of volume $\Omega = L_{x_1} \times L_{x_2} \times L_{x_3}$, where $L_{x_1} = L_{x_3} = 4\pi/\sigma$ and $L_{x_2} = 20$, with ${x_2} \in [-10,10]$.
Simulations are performed on a $512 \times 1024 \times 512$ uniform Cartesian grid which allows the developing current sheets to be resolved across multiple grid nodes. Lengths, velocities, and time are nondimensionalised with respect to $L_0 = L_{x_1}/(4\pi/\sigma)$, $u_0 = \max[\mathrm{sech}^2(x_2)]$, and $t_0 = L_0/u_0$. The Reynolds and magnetic Prandtl numbers are $\mathrm{Re} = u_0 L_0 / \nu = 3500$ and $\mathrm{Pm} = \nu / \eta = 1$, respectively, and the initial pre-dynamo Lunquist number is $S= \frac{L_0\mathit{b}}{\eta} = 52.5$. To isolate turbulent fluctuations from the evolving mean fields, a Reynolds decomposition is applied,
\begin{equation}
u_i = \overline{u_i} + u_i',
\qquad
b_i = \overline{b_i} + b_i',
\end{equation}
where overbars denote averaged quantities and primes denote fluctuations. Relevant equations become:
\begin{eqnarray}
\underbrace{\partial_t\overline{u_i}}_{\clap{\xsub{time \\ derivative}}} + \underbrace{\overline{u_j} \partial_j \overline{u_i}}_{\clap{\xsub{kinetic \\ advection}}} - \underbrace{\overline{b_j} \partial_j \overline{b_i}}_{\clap{\xsub{magnetic \\ tension}}}  = \underbrace{-\partial_i\overline{p}}_{\clap{\xsub{pressure \\ gradient}}} \nonumber\\ + \underbrace{\nu \partial_{jj} \overline{u_i}}_{\clap{\xsub{diffusion}}}  + \underbrace{\partial_j \: \overline{b^\prime_i b^\prime_j}}_{\clap{\xsub{Maxwell's \\ stresses}}} - \underbrace{\partial_j \: \overline{u^\prime_i u^\prime_j}}_{\clap{\xsub{Reynolds \\ stresses}}} - \underbrace{\frac{1}{2}\partial_i \overline{b_j b_j}}_{\clap{\xsub{magnetic \\ pressure}}} ,
\end{eqnarray}
\begin{equation}
\underbrace{\partial_t\overline{b_i}}_{\clap{\xsub{time \\ derivative}}} + \underbrace{\overline{u_j} \partial_j \overline{b_i} - \overline{b_j} \partial_j \overline{u_i}}_{\clap{\xsub{magnetic induction}}}  = \underbrace{\eta \partial_{jj} \overline{b_i}}_{\clap{\xsub{magnetic \\ diffusion}}} + \underbrace{ \partial_j \: \overline{u^\prime_i b^\prime_j}- \partial_j \: \overline{b^\prime_i u^\prime_j}}_{\clap{\xsub{turbulent electromotive \\ stresses}}} .
\end{equation}
\\
\indent Averaging is performed over an ensemble of five simulations with different realisations of $\alpha$, followed by phase averaging in the $x_1$ direction over one wavelength $2\pi/\sigma$, and pointwise averaging in the initially statistically homogeneous $x_3$ direction. This procedure isolates stochastic fluctuations while retaining coherent mean-field structures associated with the current sheets. The turbulent kinetic and magnetic energies are defined as $k = \tfrac{1}{2} \overline{u_i' u_i'}$ and $m = \tfrac{1}{2} \overline{b_i' b_i'}$, respectively. Using the Reynolds-decomposed equations, we derive evolution equations for the turbulent kinetic energy, $k$, and turbulent magnetic energy, $m$, that describe the production, dissipation, transport, and inter-domain transfer of turbulent energy. The resulting expressions read as follows:

\begin{eqnarray}
\underbrace{\partial_t{k}}_{\clap{\xsub{time \\ derivative}}} 
+
\underbrace{\overline{u_j} \partial_j k}_{\clap{\xsub{mean-field \\ advection}}} 
-
\underbrace{\overline{b_j} ( \overline{u_i^\prime \partial_j b_i^\prime})}_{\clap{\xsub{mean-field \\ transfer}}}  
=
-\underbrace{\overline{u_i^\prime \partial_i p^\prime}}_{\clap{\xsub{kinetic pressure \\ transport}}} 
+
\underbrace{\nu ( \overline{u_i^\prime \partial_{jj}  u_i^\prime})}_{\clap{\xsub{kinetic \\ dissipation}}} \nonumber \\
-
\underbrace{ \overline{u_i^\prime u_j^\prime} \partial_j{\overline{u_i}} 
+ 
\overline{u_i^\prime b_j^\prime}\partial_j{\overline{b_i}}}_{\clap{\xsub{kinetic \\ production}}}
-
\underbrace{ \overline{u_i^\prime u_j^\prime \partial_j u_i^\prime}}_{\clap{\xsub{turbulent \\ transport}}}
+
\underbrace{\overline{u_i^\prime b_j^\prime\partial_j{b_i^\prime}}}_{\clap{\xsub{turbulent \\ transfer}}} \nonumber \\
-
\underbrace{\frac{1}{2}\overline{u_i^\prime \partial_i{b_j^\prime b_j^\prime}}}_{\clap{\xsub{magnetic pressure \\ transport}}},
\label{eq:tke}
\end{eqnarray}
\begin{eqnarray}
    \underbrace{\partial_t{m}}_{\clap{\xsub{time \\ derivative}}}  
    +
    \underbrace{ \overline{u_j} (\overline{ b_i^\prime \partial_j b_i^\prime})
    - \overline{b_j} (\overline{ b_i^\prime \partial_j u_i^\prime})}_{\clap{\xsub{mean-field \\ transfer}}} 
    -
    \underbrace{\overline{b_i^\prime b_j^\prime}\partial_j\overline{u_i}
    +
    \overline{b_i^\prime u_j^\prime}\partial_j{\overline{b_i}}}_{\clap{\xsub{magnetic \\ production}}} \nonumber \\
    -
    \underbrace{\overline{b_i^\prime b_j^\prime \partial_ju_i^\prime}  
    +
    \overline{u_j^\prime b_i^\prime\partial_j b_i^\prime}}_{\clap{\xsub{turbulent \\ transfer}}}  
    =
    \underbrace{\eta (\overline{b_i^\prime \partial_{jj}b_i^\prime )}}_{\clap{\xsub{magnetic \\ dissipation}}}. \; \; \:
\label{eq:tme}
\end{eqnarray}

Equations~(\ref{eq:tke})--(\ref{eq:tme}) form the basis of our analysis, allowing us to identify the dominant mechanisms responsible for turbulent energy production during the reconnection-driven transition.
\begin{figure*}
     \centering
        \includegraphics[width=\textwidth]{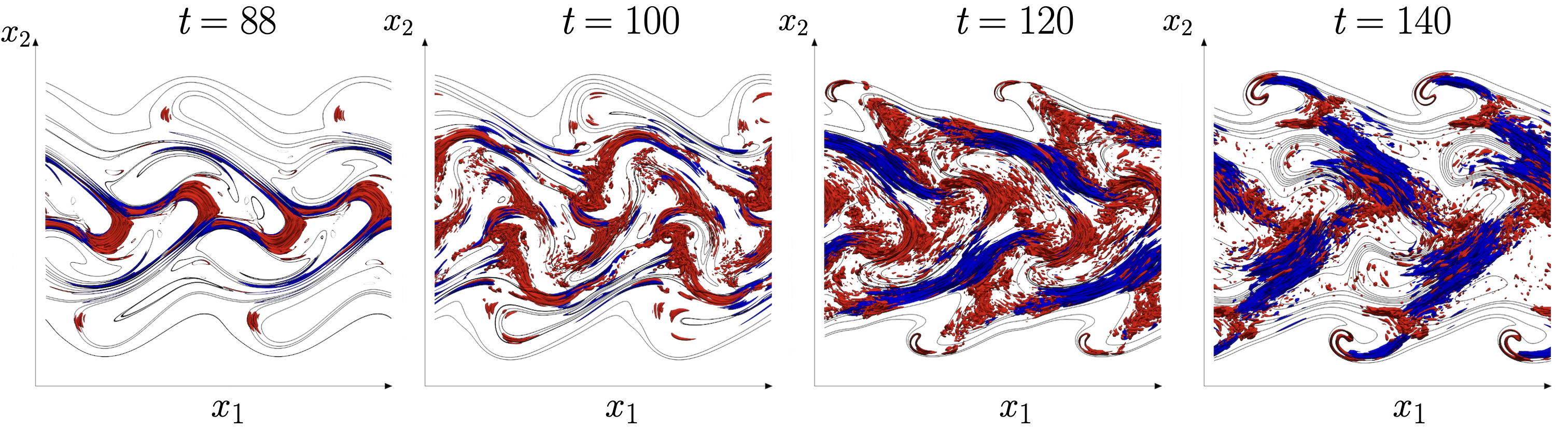}
        \caption{Planar view of the transitioning jet at representative times. Mean-field magnetic field lines (black), the 99th percentile of the current-density magnitude (blue), and the second invariant of the velocity-gradient tensor (red) are displayed.} 
        \label{fig:fig1}
\end{figure*}

\begin{figure}[!htbp]
     \centering
        \includegraphics[width=0.49\textwidth]{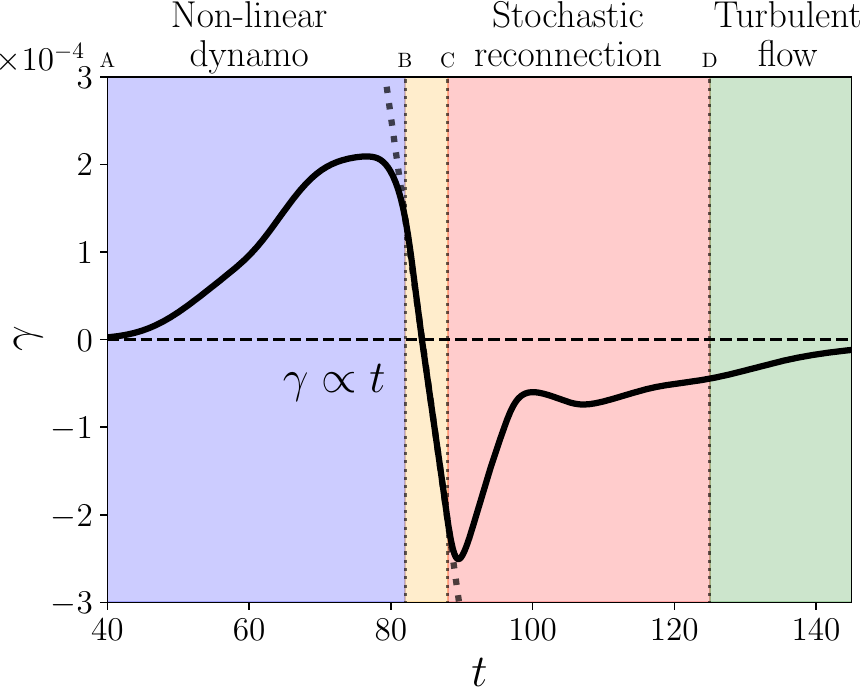}
        \caption{Time evolution of the magnetic energy growth rate $\gamma$. The linear scaling $\gamma \propto t$ is shown for reference. Distinct dynamical regimes are identified: stochastic reconnection, nonlinear dynamo, and fully developed turbulent flow. Markers A–D denote the representative time instants discussed in the text.}
        \label{fig:fig2}
\end{figure}

\begin{figure}[!htbp]
     \centering
        \includegraphics[width=0.49\textwidth]{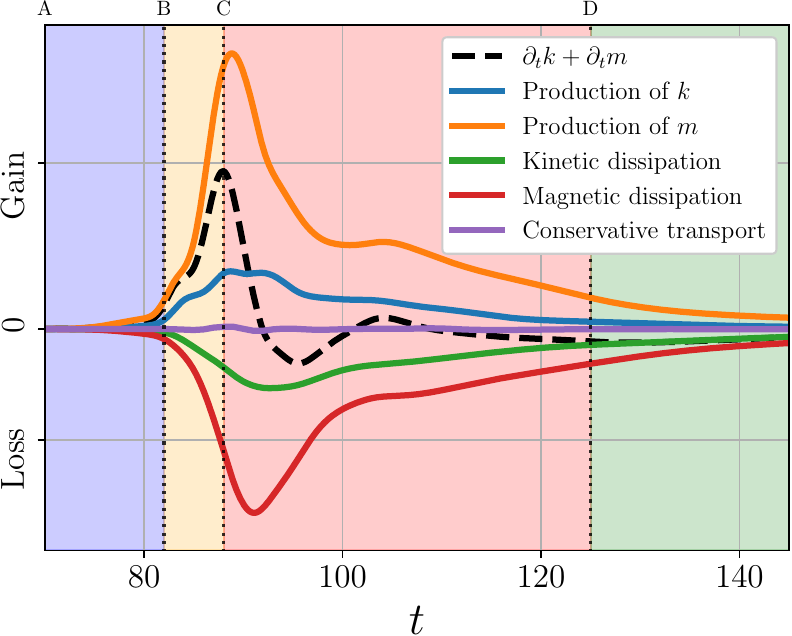}
        \caption{Turbulent energy budget across the transition and cascade. Energy fluxes entering the turbulent domain via production are treated as `gain', while fluxes removed through dissipation are treated as `loss'.}
        \label{fig:fig3}
\end{figure}

\textit{Results and discussion -} Once the simulation is initiated the hydrodynamic instability linked with the $\sigma$ mode emerges from the jet and drives the production of large vortices. This large-scale vorticity stretches and folds the initially homogenous magnetic field, facilitating kinetic to magnetic energy transfer via the mean-field dynamo.\\
\indent When the dynamo is saturated, the vortex edges begin to shear, accompanied by a gradual intensification of magnetic reconnection. A three-dimensional instability subsequently develops in the previously homogeneous $x_3$ direction, marking the onset of fully three-dimensional dynamics and the beginning of the transition to turbulence. The jet evolution is illustrated in Fig.~\ref{fig:fig1}. Current sheets rapidly form and intensify in regions of strong magnetic shear. Consistent with established theory~\cite{ruan_magnetohydrodynamic_2023}, the most intense kinetic turbulent structures develop within the reconnection outflows and persist throughout the evolution.\\
\indent To assess the transition from a dynamo-dominated to reconnection-dominated flow, we define the rate of change of the instantaneous magnetic energy integrated over the domain as
\begin{equation}
    \gamma = \partial_t \left(\frac{1}{2} \int_{\Omega} b_i b_i \, \mathrm{d}\Omega \right). 
\end{equation}
This quantity is plotted in Fig.~\ref{fig:fig2}, where different regimes are detailed below.
\begin{itemize}
    \item A-B ($t=0-82$) - Pre-turbulent growth of the non-linear dynamo.
    \item B-C ($t=82-88$) - Rapid emergence of the current sheet instability as $\gamma \propto t$.
    \item C-D ($t=88-125$) - Stochastic reconnection period.
    \item D ($t>125$) - Decaying turbulent flow. We identify the transition from stochastic reconnection to turbulence as the point at which the turbulent energy production terms settle into a stable hierarchy, marking the dominance of turbulent over mean-field dynamics.
\end{itemize}
The effects of the mean-field dynamo are initially prominent, with $\gamma$ increasing until saturation at $t=80$, followed by the onset of fast reconnection at $t=82$. During the subsequent development of the SGTR instability ($82\leq t \leq 88$), $\gamma$ grows linearly in time, $\gamma \propto t$, indicating a sustained rapid-growth phase. The evolution of the SGTR instabilities resembles findings by Oishi \emph{et al.}~\cite{oishi_self-generated_2015} and Beg \emph{et al.}~\cite{beg_evolution_2022}. Despite a modest Lundquist number which peaks at $S\approx2.2 \times10^3$, we observe a rapid growth in the current sheet instability during the SGTR phase ($82\leq t \leq 88$). This result demonstrates that current sheets can become unstable in 3D topologies independently of the plasmoid instability, which finds agreement with established theory \cite{oishi_self-generated_2015}.  We emphasise that the instability does not arise from numerical effects, as removing the seeded noise from the initial condition suppresses its development.
\\
\indent To elucidate the link between the emergent current-sheet instability and the transition to turbulence, we present the turbulent energy budget in Fig.~\ref{fig:fig3}. The production of turbulent magnetic energy is $\overline{b_i^\prime b_j^\prime}\partial_j\overline{u_i} - \overline{b_i^\prime u_j^\prime}\partial_j\overline{b_i}$, where the two terms are the interactions of the Maxwell's stresses with the mean velocity shear, and the magnetic-driven component of the turbulent electromotive stress tensor with the mean magnetic shear, respectively. Similarly the production of turbulent kinetic energy is $-\overline{u_i^\prime u_j^\prime} \partial_j\overline{u_i} + \overline{u_i^\prime b_j^\prime}\partial_j\overline{b_i}$, with the well-known Reynold's stress interacting with the velocity shear, and the kinetic-driven component of the turbulent electromotive stress tensor with the magnetic mean shear.
\begin{figure}[!htbp]
     \centering
        \includegraphics[width=0.49\textwidth]{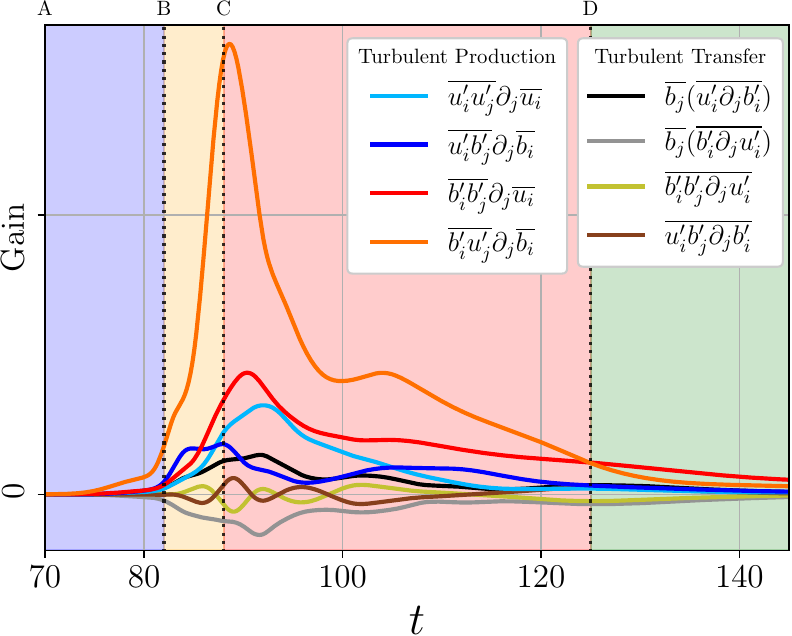}
        \caption{Dominant energy fluxes in turbulent production and transfer: EMF–magnetic-mean-shear drives production, and the resulting energy cascades between domains; dissipation and minor transport are omitted for clarity.}
        \label{fig:fig4}
\end{figure}
\\
\indent The current-sheet instability generates strong magnetic fluctuations in regions of intense mean shear. Their interaction through the EMF–shear work term $\overline{b^\prime_i u^\prime_j} \partial_j \overline{b_i}$ dominates the turbulent energy budget in Fig.~\ref{fig:fig4}, establishing a magnetically driven production mechanism responsible for the self-generation of turbulence. This finding corroborates the LV99 picture of stochastic reconnection~\cite{lazarian_reconnection_1999}, and it further clarifies the energetic pathway linking mean-field magnetic energy to the turbulent cascade.\\
\indent Although the primary route for the \emph{self-generation} mechanism and bulk production of turbulent energy may seem clear from the budgets in Figs.~\ref{fig:fig3}-\ref{fig:fig4}, we find the turbulent energy distribution during transition is far from stable. Once turbulent magnetic energy is produced, it is either dissipated via turbulent magnetic dissipation, conservatively advected, or cascaded through inter-domain transfer via both mean-field and turbulent interactions. Whilst production and dissipation terms act as sources and sinks of turbulent energy respectively, the inter-domain transfer is a non-negligible, conservative two-way process where turbulent energy is not created or destroyed, but moved between the domains of \textit{k} and \textit{m}.\\
\indent During the reconnection process, the bulk of turbulent magnetic energy is produced as Alfvén waves by the aforementioned turbulent EMF interaction with the magnetic mean-shear. Once the flow has transitioned to turbulence, the Maxwell's stress interaction with the velocity shear takes over as the primary source of Alfvénic fluctuations throughout the subsequent decaying turbulent cascade.\\
\indent In contrast to \textit{m}, the production of \textit{k} exhibits no single dominant mechanism: magnetic mean-shear and velocity mean-shear contributions alternate in prominence throughout the transition. The conservative transfer from \textit{m} to \textit{k} remains significant; at certain stages, turbulent kinetic energy gained through magnetic-fluctuation transfer (\textit{i.e.}, the conversion of Alfvénic fluctuations into kinetic fluctuations) exceeds that produced by magnetic mean-shear.
\begin{figure*}[!htbp]
    \centering
        \includegraphics[width=0.99\textwidth]{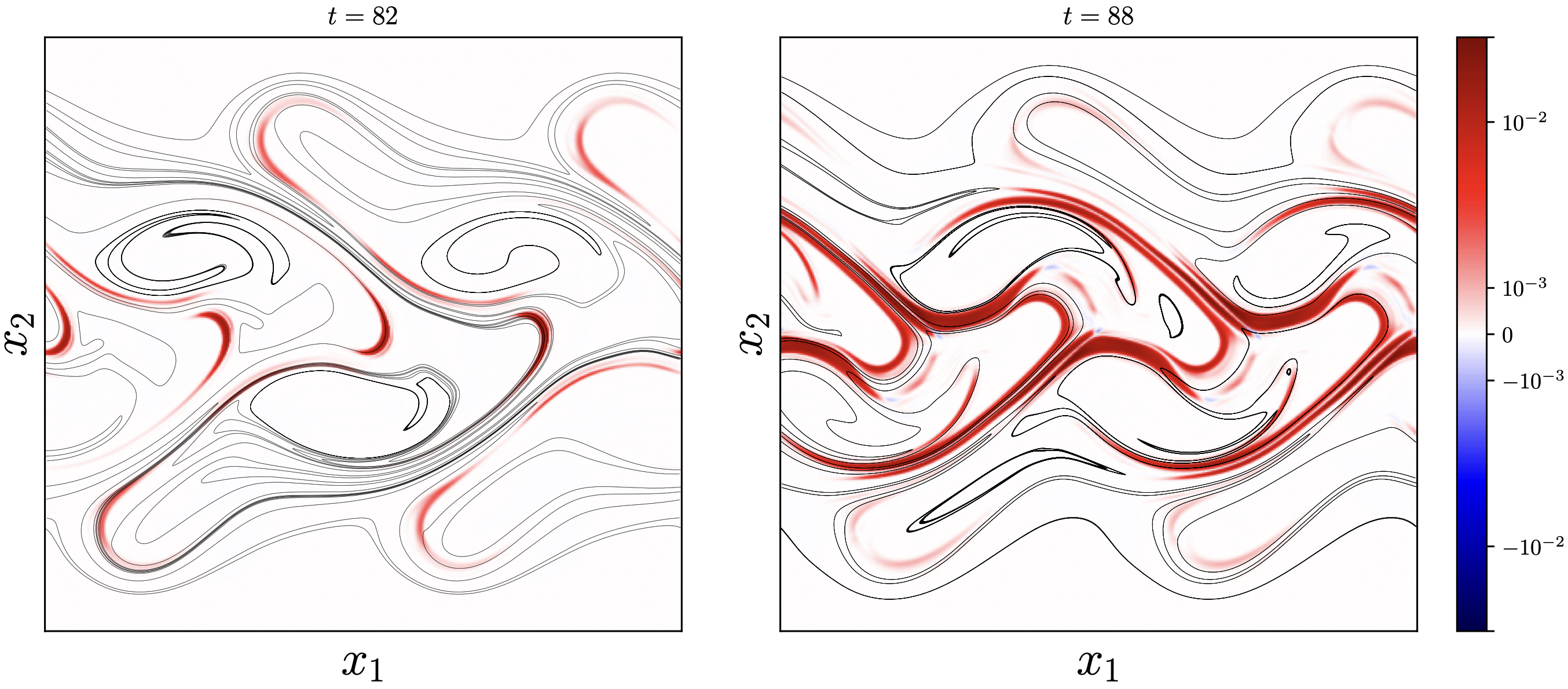}
        \caption{Magnetic field topology and turbulent energy production. Planar view of $\mathcal{E} \cdot \overline{\mathbf{J}}$ (shading) and mean magnetic field lines (contours) at $t=82$ and $t=88$. The figure illustrates the self-generation of turbulent energy within reconnection sites, driving the growth of the SGTR instability. Note the emergence of sink terms (blue) during the advanced stages of reconnection at $t=88$.}
        \label{fig:fig5}
\end{figure*}\\
\indent Regarding the locality of turbulence production, we evaluate the scalar $\mathcal{E}\cdot\overline{\mathbf{J}}$, where $\mathcal{E}=\epsilon_{kij} \langle u_i^\prime b_j^\prime\rangle$ is the turbulent EMF, $\epsilon_{kij}$ is the Levi-Civita tensor, and $\overline{\mathbf{J}}=\epsilon_{kij}\partial_i \overline{b_j}$ is the mean current density. This term quantifies the exchange of energy from the magnetic mean-field into the turbulent cascades and is illustrated in Fig.~\ref{fig:fig5} for time instants B \& C. Turbulence is strongly localised to the reconnecting current sheets, initially arising in the Alfvénic upstream and downstream regions of the X-lines where the magnetic field reverses, and subsequently spreading into the reconnection outflows. As the SGTR instability grows, $\mathcal{E}\cdot\overline{\mathbf{J}}$ increasingly fills the X-line surroundings, driving the transition from Sweet-Parker reconnection~\cite{parker_sweets_1957, sweet_neutral_1958} to stochastic reconnection~\cite{lazarian_reconnection_1999} and generating further turbulent energy. This localisation shows that the turbulent EMF is largely aligned with the mean current, acting as the main driver of the turbulent cascade. Notably, negative $\mathcal{E}\cdot\overline{\mathbf{J}}$ appears in the outflows, indicating local anti-alignment and a transfer of energy from the fluctuations back to the mean magnetic field.
\begin{figure}[!htbp]
     \centering
        \includegraphics[width=0.49\textwidth]{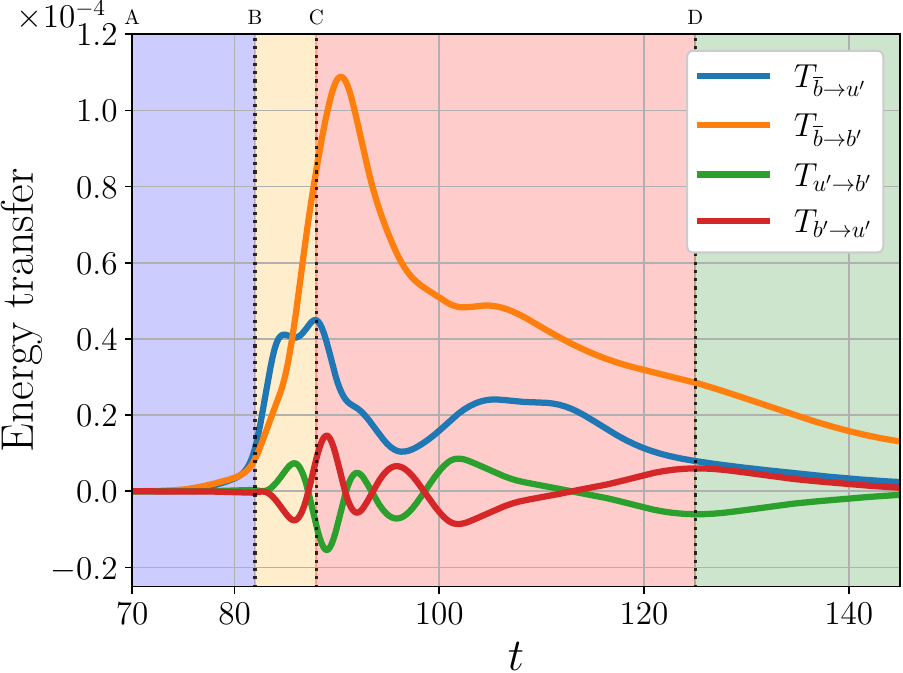}
        \caption{Energy transfer from the magnetic mean-field to Alfvénic and velocity fluctuations throughout the turbulence transition. Inter-domain energy transfer between Alfvénic and kinetic fluctuations is also shown.}
        \label{fig:fig6}
\end{figure}
\\
\indent We further define energy transfer functions derived from the mean-field equations, similar to those of Alexakis \textit{et al.}~\cite{alexakis_shell--shell_2005}, as
\begin{eqnarray}
    T_{\overline{b} \rightarrow u^\prime} = - \langle u^{\prime} \cdot (b^{\prime} \cdot \nabla) \overline{b} \rangle, \\
    T_{\overline{b} \rightarrow b^\prime } = \langle  b^{\prime} \cdot (b^{\prime} \cdot \nabla) \overline{u} \rangle, \\
    T_{ u^\prime \rightarrow b^\prime } = -\langle b^{\prime} \cdot ( b^{\prime} \cdot \nabla) u^{\prime} \rangle, \\ 
    T_{b^\prime \rightarrow u^\prime } = \langle u^{\prime} \cdot ( b^{\prime} \cdot \nabla) b^{\prime} \rangle,
\end{eqnarray}
where $\langle \cdot\rangle$ denotes the spatial mean over the computational domain. The results shown in Fig.~\ref{fig:fig6} complement our energy-budget analysis of reconnection, revealing that the majority of energy from the magnetic mean field is transferred to Alfvénic fluctuations. This energy transfer persists throughout the stochastic reconnection phase and into the fully turbulent regime, long after the SGTR has been disrupted. Another notable observation is the near-conservative balance of inter-domain energy exchange between fluctuations, which rapidly alternates during the peak reconnection activity. Whether the frequency of these alternations is specific to the reconnection geometry and time scales in our simulations remains an open question for future investigation. 

\textit{Summary -} This Letter demonstrates that three-dimensional instabilities during magnetic reconnection can self-generate sufficient turbulent energy to fully disrupt current sheets. Direct numerical simulations reveal that Alfvénic fluctuations, produced via the turbulent electromotive interaction with the mean magnetic shear, drive a cyclical amplification of turbulence that transforms laminar reconnection regions into fully turbulent ones, extending and generalising earlier observations~\cite{huang_turbulent_2016, oishi_self-generated_2015, beg_evolution_2022}. To the best of our knowledge, this is the first attempt to quantify these reconnection-driven turbulence transition mechanisms in non-stationary 3D geometries.\\
\indent We further show that the magnetic energy annihilation rate grows linearly during the SGTR growth phase, a direct consequence of the corresponding linear expansion of the effective current-sheet thickness, highlighting a robust mechanism for enhanced reconnection efficiency.\\
\indent Finally, turbulence production is spatially concentrated in reconnection regions and intensifies as the SGTR instability develops, suggesting that similar self-generated turbulence may arise in diverse reconnection scenarios and plasma regimes, including compressible systems prone to tearing-mode instabilities.

\textit{Acknowledgments -} The authors would like to thank the EPSRC for the computational time made available on the UK supercomputing     facility ARCHER/ARCHER2 via the UK Turbulence Consortium (EP/R029326/1).

\bibliographystyle{apsrev4-2}
\bibliography{References_PRL_jet}

\end{document}